\documentclass[prl,showpacs,superscriptaddress,twocolumn]{revtex4}
\usepackage{amssymb}
\usepackage{amsfonts}
\usepackage{graphicx}
\usepackage{color}

\newcommand{\be}{\begin{equation}}
\newcommand{\ee}{\end{equation}}
\newcommand{\bea}{\begin{eqnarray}}
\newcommand{\eea}{\end{eqnarray}}

\begin{document}

\title{Spin-charge disparity of polarons in organic ferromagnets}
\author{G.~C.~Hu}
\email[Electronic mail: ]{hgc@sdnu.edu.cn} \affiliation{College of Physics and
Electronics, Shandong Normal University, Jinan 250014, China}
\affiliation{Institute of Theoretical Physics, Technische Universit\"at Dresden,
01062 Dresden, Germany}
\author {H.~Wang}
\affiliation{College of Physics and Electronics, Shandong Normal University,
Jinan 250014, China}
\author {J.~F.~Ren}
\affiliation{College of Physics and Electronics, Shandong Normal University,
Jinan 250014, China}
\author{S.~J.~Xie}
\affiliation{School of Physics, Shandong University, Jinan 250100, China}
\author {C.~Timm}
\email[Electronic mail: ]{carsten.timm@tu-dresden.de}
\affiliation{Institute of Theoretical Physics, Technische Universit\"at Dresden,
01062 Dresden, Germany}

\date{May 28, 2013}

\begin{abstract}
Polaron formation in quasi-one-dimensional organic ferromagnets is
studied based on an extended Su-Schrieffer-Heeger model combined
with a Kondo term. The charge distribution of the
polaron is found to be highly asymmetric under spatial reflection,
due to the spin radicals. On the contrary, the spin density is
nearly symmetric; the spin asymmetry introduced by the extra
electron inducing the polaron formation is nearly compensated by
the spin polarization of the lower-energy states. We discuss these results
on the basis of real-space mean-field calculations and symmetry arguments.
\end{abstract}


\maketitle

\emph{Introduction.}---Organic spintronics, i.e., the manipulation
of electronic spin degrees of freedom in organic molecules, has
attracted increasing interest in recent years
\cite{Nab07205,Ded09707,Sug091738}. One of the characteristic
properties of organic materials is the strong electron-lattice
interaction. This makes the ground state of conjugated organic
materials dimerized. For example, \textit{trans}-polyacetylene,
\textit{trans}-$(\mathrm{CH})_x$, with uniform C-C bond lengths
would be a one-dimensional metal with a half-filled band. However,
the system can lower its energy by spontaneously forming a
dimerized state with alternating short and long bonds
\cite{Pei55,Hee88781}. This Peierls instability opens a
dimerization gap in the electronic spectrum, which induces the
conjugated organic materials as organic semiconductors. The strong
electron-lattice interaction also leads to another important
effect: Electrons or holes doped into a polymer typically
self-trap and form nonlinear excitations such as polarons and
bipolarons \cite{Hee88781}. A polaron consists of an electron
(hole) with specific spin together with a localized distortion of
the polymer chain. The whole polaron can move under an electric
field as a quasiparticle, which is crucial for organic
light-emitting diodes and organic spin-valve devices
\cite{Rak991518,Ren05074503}.

Organic ferromagnets are particularly fascinating since they combine
ferromagnetic and polaronic properties. Purely organic ferromagnets not
containing transition-metal ions can be realized by using spin radicals
\cite{Kor87370,Cao88625,Iwa85379,Kat91223,Sug10180401}, which are
usually heterocycles containing an unpaired electron
\cite{Alb0966}. For example, the organic ferromagnet
\emph{poly}-BIPO can be obtained from polyacetylene by replacing every other
H atom by a spin radical \cite{Kor87370,Cao88625}. There is an exchange coupling
between the $\pi$ electrons in the main carbon chain and the residual spins of
radicals. In the ground state, the radical spins order
ferromagnetically \cite{Kor87370,Cao88625,Fan943916,Fan951304}.
Recent progress in molecular electronics offers the prospect of
designing spintronics devices based on organic ferromagnets, including
spin filters \cite{Hu07165321,Wan06235325}, spin diodes
\cite{Hu08234708}, and spin valves \cite{Yoo06247205,Li121261}.
This requires to understand the property of carriers in such materials, in which
the spin coupling with radicals and the electron-lattice interaction will be
important.

In this paper, we study the effects of adding an extra electron to a
ferromagnetic polymer such as \emph{poly}-BIPO. The extra electron forms a
polaron, which is dramatically different from a polaron in a nonmagnetic chain.
Specifically, its charge density is highly asymmetric, whereas its spin
density, relative to the chain without polaron, is nearly symmetric.
We show that this disparity is enhanced by the proximity of \emph{poly}-BIPO to
the transition to a periodic phase and discuss it in terms of the ease with
which the charge and spin densities can adapt in order to lower the polaron
energy.

\emph{Model and method.}---We describe a quasi-one-dimensional
organic ferromagnet such as \emph{poly}-BIPO. The molecule
consists of a $\pi$-conjugated carbon chain and spin radicals
attached to the odd sites. Each radical contains an uncompensated
spin $\mathbf{S}_{i}$. We assume an antiferromagnetic coupling
between the spin of the $\pi$ electrons and the radical spins
$\mathbf{S}_{i}$. Such molecules can be well described by an
extended Su-Schrieffer-Heeger (SSH) model \cite{Su791698} combined
with a Kondo term. The Hamiltonian is written
as \cite{Fan943916,Fan951304}
\begin{eqnarray}
H & = & -\sum_{i,\sigma}(t_{0}-\alpha
  y_{i})(c_{i+1,\sigma}^\dagger
  c_{i,\sigma}+\mathrm{H.c.})+\frac{K}{2}\sum_{i}y_{i}^{2} \nonumber \\
&& {}+ U \sum_{i}c_{i,\uparrow}^\dagger c_{i,\uparrow}c_{i,\downarrow}^\dagger
  c_{i,\downarrow}
  + J \sum_i \delta_{i\text{ odd}}\: \mathbf{S}_{i}\cdot\mathbf{s}_{i}.
\label{SSH}
\end{eqnarray}
The first term describes the hopping of $\pi$ electrons along the
carbon chain. Here, $c_{i,\sigma}^\dagger$ ($c_{i,\sigma}$)
denotes the creation (annihilation) operator of an electron at
site $i$ with spin $\sigma$. $t_{0}$ is the hopping integral
between two neighboring sites of a uniform chain and $\alpha$
denotes the electron-lattice coupling constant. $y_{i}$ is the
lattice distortion $y_{i}\equiv u_{i+1}-u_{i}$, where $u_{i}$ is
the displacement of the carbon atom at site $i$. The second term
is the elastic energy due to the lattice distortion with elastic
coefficient $K$. The third term is the on-site electron-electron
interaction of strength $U$. The last term denotes the
antiferromagnetic coupling between the $\pi$-electron spins
$\mathbf{s}_{i} = \sum_{\sigma\sigma'} c_{i,\sigma}^\dagger
(\mbox{\boldmath$\sigma$}_{\sigma\sigma'}/2) c_{i,\sigma'}$ and
the radical spins $\mathbf{S}_{i}$, with strength $J>0$, and
$\delta_{i\text{ odd}}=1$ ($\delta_{i\text{ odd}}=0$) for $i$ odd
(even). Periodic boundary conditions are assumed.

The ground state of the molecule is calculated using a mean-field approximation
for the spin-spin and electron-electron interactions. We obtain the
eigenenergies
$\varepsilon_{\mu,\sigma}$ and the eigenstates $|\psi_{\mu,\sigma}\rangle$ with
(real) eigenfunctions $\psi_{\mu,\sigma,i}$ in Wannier space by solving the
Schr\"odinger equation
\begin{eqnarray}
\lefteqn{ \varepsilon_{\mu,\sigma}\, \psi_{\mu,\sigma,i}
  = \sum_j H^{\text{MF}}_{\sigma,ij}\, \psi_{\mu,\sigma,j} } \nonumber \\
&& \equiv -(t_{0}-\alpha y_{i})\,
\psi_{\mu,\sigma,i+1}-(t_{0}-\alpha
  y_{i-1})\, \psi_{\mu,\sigma,i-1} \nonumber \\
&& {}+ U\, \bar{n}_{i,-\sigma}\psi_{\mu,\sigma,i}
  + J\, \delta_{i\text{ odd}}\: \langle S_{i}^{z}\rangle\,
  \frac{\sigma}{2}\, \psi_{\mu,\sigma,i} , \hspace{6em}
\label{MFSchroe}
\end{eqnarray}
where $\langle S_{i}^{z}\rangle$ is the average value of the
radical spin, assumed to be in the \emph{z} direction, and
$H^{\text{MF}}_{\sigma,ij}$ is the matrix element of the
mean-field Hamiltonian for the $\pi$ electrons with spin $\sigma$.
The spin quantum number $\sigma$ assumes the numerical values
${\uparrow} \equiv 1$ and ${\downarrow} \equiv -1$. Throughout the
paper, we enumerate the states such that
$\varepsilon_{\mu',\sigma}>\varepsilon_{\mu,\sigma}$ for
$\mu'>\mu$. The average occupation number of $\pi$ electrons at
site \emph{i} with spin $\sigma$ is $\bar{n}_{i,\sigma} =
\sum_{\mu\text{ occ.}} |\psi_{\mu,\sigma,i}|^2$. The sum is over
all occupied states, i.e., all states with
$\varepsilon_{\mu,\sigma}$ up to the Fermi energy. The lattice
distortion $y_{i}=u_{i+1}-u_{i}$ in Eq.\ (\ref{MFSchroe}) is
determined by minimizing the total energy
\begin{equation}
E(\{y_{i}\}) =
  \sum_{\mu,\sigma\text{ occ.}} \varepsilon_{\mu,\sigma}(\{y_{i}\})
  + \frac{K}{2} \sum_{i} y_{i}^{2}
\label{Etotal}
\end{equation}
with respect to the distortions $y_i$. The sum $\sum_{\mu,\sigma\text{ occ.}}$
is over all occupied states. The contraint $\sum_i y_i=0$ is implemented using a
Lagrange multiplier. This leads to the equation
\begin{eqnarray}
y_i & = & -\frac{2\alpha}{K} \sum_{\mu,\sigma\;\mathrm{occ.}}
  \psi_{\mu,\sigma,i} \psi_{\mu,\sigma,i+1} \nonumber \\
&& {}+ \frac{2\alpha}{NK} \sum_k \! \sum_{\mu,\sigma\;\mathrm{occ.}}
  \psi_{\mu,\sigma,k} \psi_{\mu,\sigma,k+1} .
\label{yi_explicit}
\end{eqnarray}
Equations (\ref{MFSchroe}) and (\ref{yi_explicit}) are solved self-consistently
\cite{Fan943916}.

For the numerical calculations we use parameter values appropriate for
\emph{poly}-BIPO \cite{Fan943916,Xie00635,Hu07165321}: $t_{0}=2.5$ eV,
$\alpha=4.1$ eV/\AA, $K=21.0$ eV/\AA$^{2}$ and
$\langle S_{i}^{z}\rangle={1}/{2}$. We introduce
dimensionless interaction strengths $j=J /t_{0}$ and $u=U/t_{0}$. The
net charge density $\rho_{c,i}$ in units of the elementary charge $e>0$
and the spin density $\rho_{s,i}$ of
$\pi$ electrons in the main chain, in units of $\hbar$, are defined as
$\rho_{c,i} \equiv -(\bar{n}_{i,\uparrow}+\bar{n}_{i,\downarrow}-1)$ and
$\rho_{s,i} \equiv (\bar{n}_{i,\uparrow}-\bar{n}_{i,\downarrow})/2$,
respectively.

\emph{Neutral state.}---Here, we consider a half-filled
chain with $N=100$ sites. To focus on the effect of the spin radicals, we start
with the case without electron-electron interactions, $u=0$.
Without spin radicals ($j=0$), the model reverts
to a normal polymer. In the ground state the carbon
atoms undergo a dimerization \cite{Pei55,Hee88781}
and a large energy gap (approximately $1.4\,\mathrm{eV}$) opens up.

In the presence of spin radicals ($j>0$), the radical spins order
ferromagnetically due to their coupling to the $\pi$ electrons. As
shown in previous numerical studies \cite{Fan943916,Xie00635}, a
spin density wave (SDW) is formed along the main chain, while the
net charge density remains zero. This can be understood as
follows: For fixed distortions $y_i$, the states
$|\psi_{\mu,\sigma}\rangle$ are eigenstates of the mean-field
Hamiltonian with eigenenergies $\varepsilon_{\mu,\sigma}$ and of
the spin $z$ component with eigenvalues $\sigma/2$. We construct
new states $|\psi_{\mu',-\sigma}\rangle$ by rotating the spin
state by $180^\circ$ about the $x$ or $y$ axis and multiplying the
wave function by $(-1)^i$, i.e., $\psi_{\mu',-\sigma,i} = (-1)^i\,
\psi_{\mu,\sigma,i}$. The mean-field Schr\"odinger equation
(\ref{MFSchroe}) with $U=0$ shows that
$|\psi_{\mu',-\sigma}\rangle$ is an eigenstate with eigenenergy
$\varepsilon_{\mu',-\sigma} = -\varepsilon_{\mu,\sigma}$.
According to our enumeration of states, we thus have $\mu' =
N+1-\mu$. The charge density of the states
$|\psi_{\mu',-\sigma}\rangle$ and $|\psi_{\mu,\sigma}\rangle$ is
clearly identical. The completeness relation ensures that the
total charge density would be uniform, regardless of the
distortions, if \emph{all} states were occupied. In the neutral
state, the chain is half filled. Thus, exactly one of the two
states $|\psi_{\mu',-\sigma}\rangle$ and
$|\psi_{\mu,\sigma}\rangle$ is occupied. Since they have the same
charge density, the total charge density of the half-filled chain
is half that of the completely occupied chain and is thus uniform.

There is no analogous symmetry argument that protects the spin density. The spin
density would vanish for a completely occupied chain, but since the states
$|\psi_{\mu',-\sigma}\rangle$ and $|\psi_{\mu,\sigma}\rangle$ have
\emph{opposite} spin density, the half-filled, neutral chain is not required to
have vanishing spin density. As expected for antiferromagnetic coupling $j>0$,
the spin density is negative at the odd sites, connected to the radical spins.
However, the \emph{total} spin in the neutral state is still zero, in spite of
the radical spin preferring electronic spin down. To understand this, consider
another state $|\psi_{\mu'',-\sigma}\rangle$ that is obtained from
$|\psi_{\mu,\sigma}\rangle$ by rotating the spin state
by $180^\circ$ about the $x$ or $y$ axis and taking the mirror
image of the wave function with respect to a bond center.
This transformation maps short (long) bonds onto short (long) bonds but odd
(even) sites onto even (odd) sites. We assume that the displacement $u_i$
is symmetric with respect
to the mirror plane, which is the case for the periodic dimerized chain.
Then it is straightforward to show using Eq.\ (\ref{MFSchroe}) with $U=0$ that
$|\psi_{\mu'',-\sigma}\rangle$ is also an eigenstate but with the eigenenergy
$\varepsilon_{\mu'',-\sigma} = \varepsilon_{\mu,\sigma} - \sigma J/4$. The index
is $\mu'' = \mu$ since the mapping does not change the order of states.
We see that for $J=0$ (no radicals), all eigenstates form degenerate pairs of
opposite spin. For any even filling, the total spin vanishes. A nonzero exchange
interaction $J>0$ lifts the degeneracy. However, the Fermi energy for the
neutral state lies in the dimerization gap. As long as the
energy splitting $J/2$ is smaller than the dimerization gap, the states
$|\psi_{\mu,\sigma}\rangle$ and $|\psi_{\mu'',-\sigma}\rangle$ are still filled
in pairs and the total spin remains zero. For the parameters used here, this is
satisfied.

Like in nonmagnetic polymers such as
\textit{trans}-$(\mathrm{CH})_x$, the short bonds can be either
the even or the odd ones---the dimerization spontaneously breaks
translation symmetry by doubling the unit cell, leading to two
degenerate ground states. This is still the case for the magnetic
polymer. However, now the dimerization in addition breaks
reflection symmetry, the short bonds can be either to the left or
to the right of a radical spin and these two ground states are not
symmetric under spatial reflection.

\emph{Polaron state.}---We now consider the state with one extra electron.
From the previous discussion, one would expect it to occupy the lowest
unoccupied molecular orbital (LUMO) with spin down. This is indeed found.

\begin{figure}[tb]
\includegraphics[width=\columnwidth]{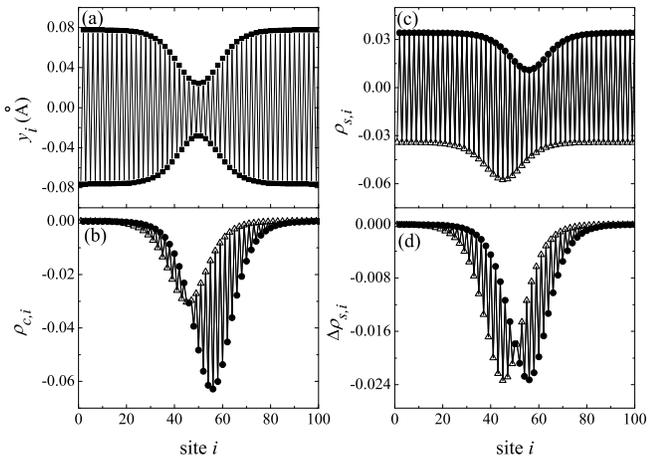}
\caption{Polaron state of the organic ferromagnet for exchange
coupling $j=0.5$: (a) Lattice distortion, (b) net charge density,
(c) spin density, and (d) net spin density of the polaron
calculated by subtracting the spin density of the neutral state
from that of the charged one. Open triangles (filled circles)
correspond to odd (even) sites. The same symbols are also used in
the following figures.} \label{fig1}
\end{figure}

The exchange interaction is set to
$j=0.5$. The lattice distortion, the net charge density, and the spin density
are shown in Figs.\ \ref{fig1}(a)--\ref{fig1}(c), respectively. It is clearly
seen that the extra electron leads to the formation of a polaron.
The distortion $y_i$ is reduced within the polaron---the
dimerization is weaker there---and is symmetric with respect
to the center of the polaron, within numerical accuracy.
This is similar to the case of a normal polymer \cite{SuS80}.
Additional insight can be gained by considering the \emph{net spin density}
$\Delta\rho_{s,i}\equiv\rho_{s,i} - \rho_{s,i}|_{\text{neutral}}$
plotted in Fig.\ \ref{fig1}(d),
obtained by subtracting the spin density of the neutral
state from the one of the charged state.
Compared to polarons in nonmagnetic polymers, two distinctive
effects are observed: (\textit{i}) The charge
distribution is asymmetric with respect to the center of the polaron
(the middle of the bond between sides 50 and 51 in the present case).
Figure \ref{fig1}(b) shows that net charge density on both the odd and the even
sites is peaked, but with peak positions shifted to the left and right,
respectively. Furthermore, the peak for the even sites is much larger than
the one for the odd sites. (\textit{ii}) The net (or excess) spin density
does not coincide with the highly asymmetric charge density but instead is
nearly symmetric; close examination of the data shows that there is a small
asymmetry. Evidently the spin is distributed differently from the charge.

\begin{figure}[tb]
\includegraphics[width=\columnwidth]{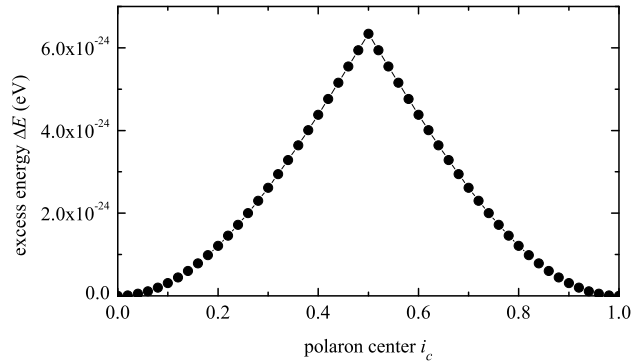}
\caption{Ground-state energy relative to its minimum of a polaron
with constrained center $i_c$, as a function of $i_c$. Here,
$i_c=0$ and $1$ correspond to the bond-centered configuration,
whereas $i_c=1/2$ corresponds to the site-centered configuration.
} \label{fig2}
\end{figure}

Moreover, in the ground state, the polaron is centered at a bond
and the distortions are symmetric with respect to this center, see
Fig.\ \ref{fig1}(a). This result is not trivial to obtain
numerically because the energy differences involved are tiny.
Figure \ref{fig2} shows the energy as a function of the polaron
center $i_c$. To obtain this quantity, we have minimized the total
energy, Eq.\ (\ref{Etotal}), under the additional constraint of a
fixed polaron center $i_c$. We write this constraint as
\begin{equation}
C\, \sum_k (-1)^k\, y_k \left(\begin{array}{c}
    \cos\frac{2\pi k}{N} \\[0.6ex]
    \sin\frac{2\pi k}{N}
    \end{array}\right)
  = -\left(\begin{array}{c}
    \cos\frac{2\pi i_c}{N} \\[0.6ex]
    \sin\frac{2\pi i_c}{N}
    \end{array}\right) ,
\label{center_constraint}
\end{equation}
where
$C$ is a normalization factor. This definition of $i_c$ implements
the periodic boundary conditions by mapping the chain onto a ring.
The factors $(-1)^k$ take care of the fact that the distortions
are staggered, while the additional factor of $-1$ on the
right-hand side is needed since the distortions are \emph{reduced}
in the region of the polaron so that the center of mass of $|y_k|$
is at the antipodal point of the polaron center. We implement Eq.\
(\ref{center_constraint}) with additional Lagrange multipliers.
Since the energy difference between polaron states with different
$i_c$ are so small, we used Mathematica \cite{Mathematica} with
precision set to 32 digits to obtain Fig.\ \ref{fig2}. The energy
differences are small because the polaron is relatively large,
i.e., it comprises many sites, see Fig.\ \ref{fig1}. In the limit
of diverging polaron size, a continuum model would become exact,
which of course does not have any energy barrier for displacing
the polaron.

\begin{figure}[tb]
\includegraphics[width=\columnwidth]{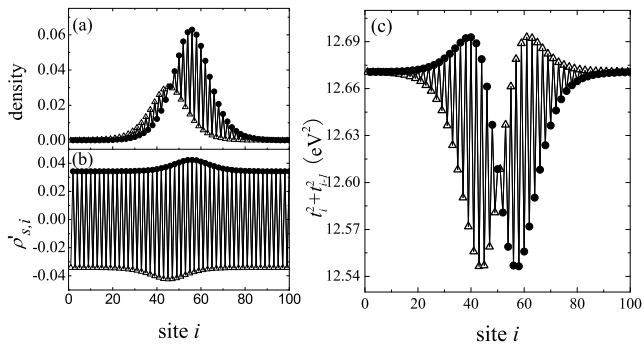}
\caption{(a) Probability density $|\psi_{\mu,\sigma,i}|^2$ of the
orbital occupied by the additional spin-down electron. (b) Spin
density $\rho'_{s,i}$ contributed by the Fermi sea in the polaron
state. (c) Sum of the square of hopping amplitudes $t_{i} =
t_{0}-\alpha y_{i}$ between site $i$ and its neighbors.}
\label{fig3}
\end{figure}

To explain the anomalous spin and charge distribution, we first examine the wave
function of the orbital $|\psi_{N/2+1,{\downarrow}}\rangle$
occupied by the extra spin-down electron.
Figure \ref{fig2}(a) shows the probability density of this orbital.
The orbital has higher weight at the even sites
compared to the odd sites, which is consistent with the
charge distribution shown in Fig.\ \ref{fig1}(b). The antiferromagnetic
coupling actually favors spin-down electrons at the \emph{odd} sites, which are
the ones coupled to the radical spins. However, the orbital is
antibonding so that its spatial distribution is opposite to
the bonding states and has higher weight at the \emph{even} sites.
Indeed, it has the same probability density as its
\emph{bonding} spin-up partner state $|\psi_{\mu',{\uparrow}}\rangle$
constructed above.

In Fig.\ \ref{fig3}(a), the maximum of the density on even (odd)
sites is shifted to the right (left) with respect to the polaron
center. This is correlated with the even bonds being long
($y_i>0$) in the ground state found here, see Fig.\ \ref{fig1}(a).
For the degenerate state with short even bonds we find the
reversed pattern.
The localization of the extra electron is due to the change in the
hopping amplitudes $t_{i}\equiv t_{0}-\alpha y_{i}$. Figure
\ref{fig3}(c) shows the quantity $t_{i}^{2}+t_{i-1}^{2}$ for each
site, which is a measure of the hopping rate of an electron
between site $i$ and its neighbors. In spite of a small increase
for some sites, this quantity is obviously reduced in the polaron
region. In particular, the largest reductions and thus the largest
tendencies toward localization occur at sites displaced to the
right (left) for even (odd) sites. This explains why the maximum
density $\psi_{N/2+1,{\downarrow},i}^2$ on even (odd) sites is
displaced to the right (left).

The spin and charge densities contributed by the orbital
containing the extra electron are of course proportional to each
other. Why, then, does the spin density of the polaron not follow
the charge density? To analyze this, we have to consider also the
other occupied states at lower energies. We here call these states
the \emph{Fermi sea}. First, we observe that the net charge
density carried by the Fermi sea vanishes (not shown). This
follows immediately from the argument given above for the uniform
charge density in the neutral state---this argument works for
arbitrary distortions $y_i$. Consequently, the charge density of
the polaron is determined entirely by the single orbital
considered above. On the other hand, the net spin density
contributed by the Fermi sea, plotted in Fig.\ \ref{fig3}(b),
clearly is affected by the polaron formation. The amplitude of the
spin density wave is enhanced both for the odd (spin-down) and for
the even (spin-up) sites but the maximum enhancements are shifted
in opposite directions relative to the polaron center. The spin
density $\rho'_{s,i}$ of the Fermi sea is odd under reflection in
a mirror plan through the polaron center, i.e., it has the
symmetry property $\rho'_{s,j}=-\rho'_{s,i}$, where $j$ is
obtained from $i$ by reflection. Evidently, the highly asymmetric
net spin densities contributed by the orbital containing the extra
electron and by the Fermi sea combine to form the nearly, but not
quite, net symmetric spin density of the polaron.

In the following, we discuss these numerical findings. First,
there is no symmetry that would require the spin density of the
Fermi sea to vanish---it does not vanish even for the neutral
state---or to be periodic for aperiodic distortions $y_i$.
However, the spin density of the Fermi sea is odd under reflection
with respect to the polaron center, see Fig.\ \ref{fig3}(b), and
its total spin thus vanishes. This can be understood based on the
second transformation discussed for the neutral case: The state
$|\psi_{\mu'',-\sigma}\rangle$ obtained from
$|\psi_{\mu,\sigma}\rangle$ by spin rotation and spatial
reflection with respect to a bond center is still an eigenstate in
the polaron case, provided that this reflection leaves the
distortions invariant. This requires the distortions to be
symmetric with respect to the polaron center and in particular the
polaron to be centered at a bond. This is the case, as we have
shown above, see Fig.\ \ref{fig2}. The corresponding eigenenergy
is $\varepsilon_{\mu'',-\sigma} = \varepsilon_{\mu,\sigma} -
\sigma J/4$. As long as the exchange splitting $J/2$ is smaller
than the dimerization gap, which is the case for our parameters,
in the Fermi sea either both or neither of the states
$|\psi_{\mu'',-\sigma}\rangle$ and $|\psi_{\mu,\sigma}\rangle$ are
occupied. The spin densities of these two states are related by
multiplication by $-1$ and spatial reflection. Their combined spin
density is then invariant under this transformation. Consequently,
the total spin density of the Fermi sea is odd under reflection.
Thus the total spin of the Fermi sea vanishes.

\begin{figure}[tb]
\includegraphics[width=\columnwidth]{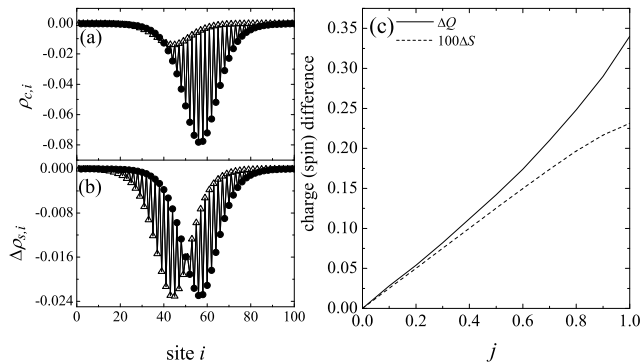}
\caption{(a) Net charge density and (b) net spin density of a
polaron with a stronger exchange coupling $j=1.0$. (c) Charge
asymmetry $\Delta Q$ (solid curve) and spin asymmetry $\Delta S$
(dashed curve) as functions of the exchange coupling $j$. The spin
difference has been multiplied by $100$ to make it more clearly
visible.} \label{fig4}
\end{figure}

Second, the question remains of why the asymmetric spin densities
of the Fermi sea and of the extra electron combine to form a
nearly symmetric net spin density, see Fig.\ \ref{fig1}(d).
Figures \ref{fig4}(a) and \ref{fig4}(b) show the net charge and
spin densities of a polaron for the stronger exchange coupling
$j=1.0$. The larger exchange coupling markedly increases the
charge asymmetry but the spin asymmetry remains small. This begs
the question whether the spin asymmetry might be smaller than the
charge asymmetry because it is of higher order in the perturbation
$j$ relative to the nonmagnetic chain. To answer this, it is
useful to quantify the spin and charge asymmetries. Fixing the
polaron center to the center of the bond between sites $N/2$ and
$N/2+1$ without loss of generality, we define the spin and charge
asymmetries as
\begin{eqnarray}
\Delta S & \equiv & \left|\sum_{i=N/2+1}^N \Delta\rho_{s,i}
  - \sum_{i=1}^{N/2} \Delta\rho_{s,i} \right| ,
  \\
\Delta Q & \equiv & \left|\sum_{i=N/2+1}^N \rho_{c,i}
  - \sum_{i=1}^{N/2} \rho_{c,i} \right| ,
\end{eqnarray}
respectively. Figure \ref{fig4}(c) shows $\Delta S$ and $\Delta Q$
as functions of the exchange coupling $j$. Clearly, both
asymmetries appear in first order in $j$. Thus an argument based
only on perturbation theory in $j$ cannot explain the small spin
asymmetry.

\begin{figure}[tb]
\includegraphics[width=\columnwidth]{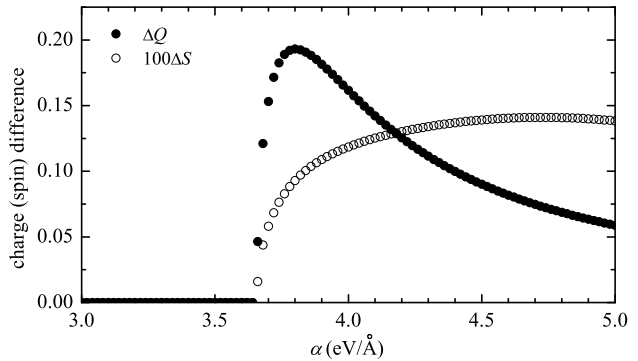}
\caption{Charge asymmetry $\Delta Q$ and spin asymmetry $\Delta S$
as functions of the electron-lattice coupling constant $\alpha$
for chain length $N=100$. The spin difference has been multiplied
by $100$ to make it more clearly visible.} \label{fig5}
\end{figure}

Hence, one might ask whether there is another small parameter.
Such a small parameter indeed exists, namely the distance of the
electron-lattice coupling $\alpha$ to its critical value
$\alpha_c$, where polaron formation sets in. Figure \ref{fig5}
shows the spin and charge asymmetries as functions of $\alpha$,
for otherwise the same parameters as above, in particular for a
chain of length $N=100$. Note that the charge asymmetry increases
strongly as the critical coupling strength $\alpha_c$ is
approached from above, whereas the spin asymmetry decreases toward
zero. For longer chains, the critical point $\alpha_c$ shifts to
slightly lower values, the charge asymmetry increases even further
(the downturn of $\Delta Q$ below $\alpha\approx
3.8\,\mathrm{eV}/\text{\AA}$ in Fig.\ \ref{fig5} is due to the
polaron size becoming comparable with the chain length $N$), and
the spin asymmetry still becomes small close to $\alpha_c$ (not
shown). However, an additional phase with two solitons instead of
one polaron intervenes between the polaron and periodic phases. At
the value of $\alpha=4.1\,\mathrm{eV}/\text{\AA}$ assumed above,
the dependence on the system size $N\ge 100$ is very weak.

The dependence on $\alpha$ shown in Fig.\ \ref{fig5} suggests that
the spin asymmetry is much smaller than the charge asymmetry due
to the vicinity of $\alpha$ for \textit{poly}-BIPO to the critical
value. Why, then, is the spin asymmetry small for
$\alpha\gtrsim\alpha_c$, whereas the charge asymmetry is large?
The following is a plausible, though not rigorous, explanation. It
is natural to expect the energy of a possible polaron
configuration to increase with both the charge and the spin
asymmetries, since if we impose a large (charge or spin) asymmetry
as an additional constraint, the distortions $y_i$ have to adapt,
which increases the elastic energy. This increase is not balanced
by any term in the Hamiltonian (\ref{SSH}) that would favor large
asymmetries. However, the polaron cannot achieve a small
\emph{charge} asymmetry because the charge asymmetry is entirely
due to the state occupied by the additional electron. Its width is
controlled by the changes of the distortions $y_i$ in the polaron
region, as shown above. For $\alpha\gtrsim\alpha_c$, the
deviations of $y_i$ and thus of the hopping amplitudes $t_{i} =
t_{0}-\alpha y_{i}$ from the periodically dimerized chain are
small. On the other hand, the exchange coupling $J$ is constant
and thus becomes \emph{strong} relative to the aperiodic part of
the $t_i$, which controls the localization. Hence, the extra
electron, being in an antibonding state, will be nearly
exclusively localized at even sites. But as discussed above, the
maxima of the density for even and odd sites are displaced with
respect to the polaron center in opposite directions, cf.\ Fig.\
\ref{fig2}(a). Thus the extra electron is mostly displaced in one
direction for $\alpha\gtrsim\alpha_c$. In addition, the transition
between the polaron state and the periodically dimerized state
appears to be continuous or at most weakly first order so that the
polaron size has to grow for $\alpha\to\alpha_c^+$. Hence, the
extra electron is not only displaced mostly in one direction, it
is also displaced by a large distance, leading to a large charge
asymmetry.

On the other hand, many states contribute to the \emph{spin} asymmetry. The
largest contributions come from occupied states close to the Fermi energy since
states deep in the Fermi sea are little affected by the polaron formation. The
number of contributing states increases with the polaron size---if the polaron
size is comparable to the chain length $N$, all states are affected. Thus for
$\alpha\to\alpha_c^+$, many states contribute. They are not subject to any
simple constraint except that the total-spin \textit{z} component is $-1/2$.
Based on our assumption that small asymmetries are energetically favorable, the
polaron is therefore able to effectively minimize its energy by minimizing the
spin asymmetry.

\begin{figure}
\includegraphics[width=3.50 in]{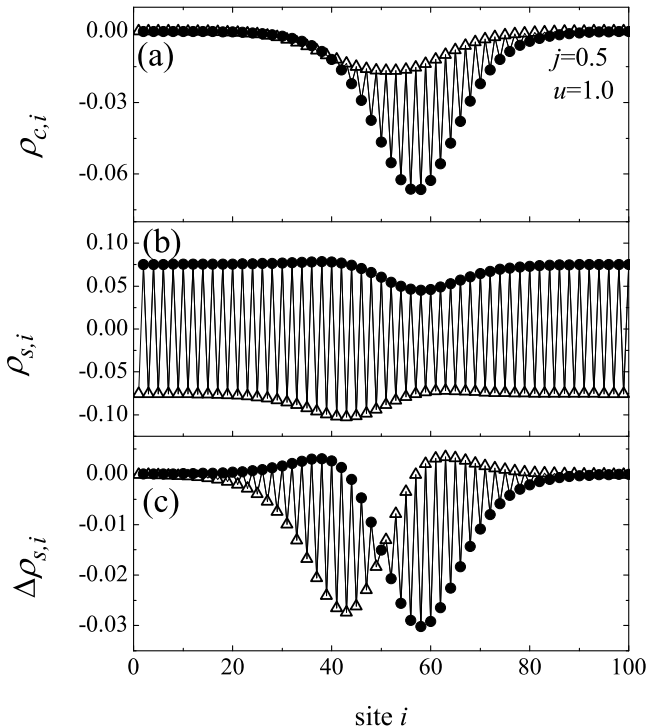}
\caption{Polaron state in the presence of electron-electron
interaction: (a) Net charge density, (b) spin density and (c) net
spin density of the polaron for $j=0.5$ and $u=1.0$.} \label{fig6}
\end{figure}

\emph{Electron-electron interaction.}---We now examine the
consequences of the electron-electron interaction. Here we set
$u=1.0$, while the other parameters are same as used for Fig.\
\ref{fig1}. The results in Fig.\ \ref{fig6} show that the picture
of spin-charge disparity is robust even in the presence of
electron-electron interaction. The effects of the
electron-electron interaction are reflected in two aspects: First,
the difference in charge density at even and odd sites is
enhanced. Second, the amplitude of the oscillation in the spin
density is increased both for the total and the net, localized
contribution. This is because the repulsion of electrons with
different spins on each site favors the formation of the SDW. It
thus acts similarly to a stronger exchange coupling $J$. Other
values for the electron-electron interaction have also been
examined and the same picture is found provided that $u$ is not
too large.

Finally, we should point out that the different behavior of spin and charge
explored here is distinct
from the well-known spin-charge separation in Luttinger liquids
\cite{Lut631154,Kim964054}, where the spin (spinon) and charge
(holon) are different collective modes of the correlated electrons
and generally move with different velocities. In our case, the observed
spin-charge disparity is mediated by the polaronic lattice
distortion, and of course both spin and charge are confined within the
polaron. The motion of such a polaron under an external field can be expected
to be nontrivial. We leave this issue to future work.

\emph{Summary.}---We have investigated the spin and charge properties of
polarons in quasi-one-dimensional organic ferromagnets with spin radicals
attached to every second site, for example \emph{poly}-BIPO. The
results show spin-charge disparity for the polarons: The charge density
localized in the polaron region is highly asymmetric, whereas the spin density
contributed by the polaron shows only very weak asymmetry. We attribute this
disparity to the observation that the charge density of the polaron cannot
efficiently adapt to minimize the polaron energy since it is carried by a
single electronic state, whereas the spin density can adapt efficiently, being
carried by many electrons. The effect is enhanced in the vicinity of the
transition between the polaron state and the periodically dimerized phase,
which is applicable for \emph{poly}-BIPO.

\emph{Acknowledgments.}---Support from the National Natural
Science Foundation of China (Grant Nos.~10904084, 10904083, and
11174181), the Shandong Province Middle-Aged and Young Scientists
Research Awards Foundation (Grant No. 2009BS01009), and the German
Science Foundation, partially through Research Unit FOR 1154
``Towards Molecular Spintronics'' are gratefully acknowledged.
Helpful discussions with Dr.\ P. M. R. Brydon and Prof.\ Jianhua
Wei are acknowledged.

\end{document}